\documentclass[twocolumn,showpacs,prd,floatfix,axodraw]{revtex4}
\usepackage{mathrsfs}
\usepackage{graphicx,,booktabs,bm}
\usepackage{overpic}
\usepackage{color}
\usepackage{amssymb}

\usepackage{mathrsfs,bm,amsmath,amssymb}
\usepackage{longtable,lscape}
\usepackage{txfonts}
\usepackage{amssymb}
\usepackage{indentfirst}
\usepackage{graphicx,,booktabs}
\usepackage{multirow,ulem}

\usepackage{color}
\usepackage{amssymb}

\definecolor{cover}{rgb}{0.77,0.87,0.88}
\definecolor{blueone}{rgb}{0.1,0.1,.7}
\definecolor{citec}{rgb}{0.14,0.47,0.09}
\definecolor{two}{rgb}{0.0,0.5,0.}
\definecolor{three}{rgb}{.5,.1,0.15}
\usepackage[bookmarks=true,bookmarksopen=false,plainpages=false,breaklinks=true,
   bookmarksnumbered=true,hypertexnames=false,
   filecolor=blue,urlcolor=three,menucolor=three,
   linkcolor=three,citecolor=blueone, colorlinks,
   anchorcolor=blue,runcolor=pink,frenchlinks=red
   pdfstartview=FitH,pdftitle=title,%
   pdfauthor=author]{hyperref}

\begin{document}
\title{Evidence for New $D_s$-Family Molecular States}
\author{Dan Jiang}
\author{Yin Huang\footnote{corresponding author}} \email{huangy2019@swjtu.edu.cn}
\author{JiongJiong Zhao}
\affiliation{School of Physical Science and Technology, Southwest Jiaotong University, Chengdu 610031,China}

\begin{abstract}
Motivated by the observed $KD^{(*)}$ molecular candidates $D_{s0}(2317)$ and $D_{s1}(2460)$, their bottom--strange counterparts, $K\bar{B}^{(*)}$ molecular states, are naturally expected, although not yet experimentally established. This discrepancy may reflect sizable heavy-quark flavor symmetry breaking, which introduces significant model uncertainties. Current studies of heavy-quark flavor symmetry breaking effects still exhibit strong parameter dependence, and further experimental input is required to constrain these effects, in particular regarding possible additional $K^{(*)}D^{(*)}$ and $K^{(*)}\bar{B}^{(*)}$ molecular states.  In this work, we examine whether additional $K^{*}D^{(*)}$ molecular states can be identified among the observed $D_s$ resonances. Within the Gaussian expansion method, we solve the Schr\"{o}dinger equation using $\sigma$, $\rho$, $\omega$, $\pi$, and $\eta$ exchange potentials, systematically including $S$-wave and higher partial waves.  We find that $D_{s1}(2700)$ can be interpreted as a pure $P$-wave $DK^{*}$ molecule, while $D_{s1}(2860)$ and $D_{s3}(2860)$ are well described as $D^{*}K^{*}$ molecular states dominated by the $^{1}P_{1}$ and $^{5}P_{3}$ components, respectively. We also predict additional molecular states with various $J^{P}$ quantum numbers.  These results provide a new description of the charmed-strange spectrum, and once confirmed will provide additional input data for studies of heavy-quark flavor symmetry breaking effects.

\end{abstract}

\date{\today}


\maketitle
\section{Introduction}\label{sec:intro}
Over the past decade, with the continuous advancement of experimental techniques, an increasing number of exotic hadronic states have been observed~\cite{ParticleDataGroup:2024cfk}.
The internal structure of these particles cannot be explained within the framework of the conventional quark model---that is, they cannot be simply interpreted as mesons composed of
a quark--antiquark pair or baryons made of three quarks.  To date, several prominent examples have been identified. In the light-quark sector, the $\Lambda(1405)$ is widely regarded
as a molecular state dominated by the $\bar{K}N$ component~\cite{Oset:1997it}. In the heavy-quark sector, the spectrum of exotic hadrons is even richer. A particularly representative
case is the $X(3872)$, first discovered by the Belle Collaboration in 2003~\cite{Belle:2003nnu}. Owing to its proximity to threshold, isospin-violating strong decays, radiative decay
patterns, and production characteristics in high-energy heavy-ion collisions, it is difficult to accommodate within the conventional meson picture.  As a result, it is broadly considered
a strong candidate for a $D\bar{D}^{*}$ molecular state~\cite{Brambilla:2019esw,Chen:2022asf,Meng:2022ozq}.  In addition, the hidden-charm pentaquark states $P_c$, observed in 2015 and 2019
by the LHCb Collaboration, are widely believed to possess multiquark configurations~\cite{Chen:2019bip,Guo:2019fdo,Xiao:2019aya,He:2019ify,Xiao:2019mvs,Roca:2015dva,Chen:2015moa,Chen:2015loa,Yang:2015bmv,Huang:2015uda,Du:2019pij}.
However, an important unresolved issue remains: apart from the experimentally observed $J/\psi\, p$ decay channel~\cite{LHCb:2015yax,LHCb:2016ztz,LHCb:2016lve,LHCb:2019kea}, why have
other decay modes predicted by theoretical models~\cite{Chen:2019bip,Guo:2019fdo,Xiao:2019aya,He:2019ify,Xiao:2019mvs,Roca:2015dva,Chen:2015moa,Chen:2015loa,Yang:2015bmv,Huang:2015uda,Du:2019pij}
not yet been observed~\cite{LHCb:2024pnt}? This puzzle has become a central topic in current particle physics research.

A more pressing issue that deserves attention is closely related to the charm--strange sector, involving the exotic states $D_{s0}(2317)$ and $D_{s1}(2460)$.  They were first observed in
isospin-violating decay channels, namely $\pi D_s$~\cite{BaBar:2003oey} and $\pi D_s^{*}$~\cite{CLEO:2003ggt,Belle:2003kup}, respectively, which already indicates their unusual nature.
Their measured masses lie significantly below quark-model expectations~\cite{Godfrey:1985xj}, leading to their widely accepted interpretation as exotic states.  At present, they are commonly
described as hadronic molecules, identified as $KD$ and
$KD^{*}$ bound states~\cite{Kolomeitsev:2003ac,Barnes:2003dj,Faessler:2007gv,Meng:2022ozq,Xie:2010zza,Guo:2006fu,Guo:2006rp,Gamermann:2006nm,Zhu:2019vnr,Mohler:2013rwa,Altenbuchinger:2013vwa}.
These features are consistent with the expectations of heavy-quark spin symmetry (HQSS)~\cite{Isgur:1991wq, Liu:2019zoy}, which emerges in the limit of infinite heavy-quark mass. In this limit,
the heavy quark acts as a static color source and its spin decouples from the light degrees of freedom, so that the total angular momentum of the light components is conserved. As a result,
$D_{s0}(2317)$ and $D_{s1}(2460)$ can be organized into a heavy-quark spin doublet, differing only in the spin configuration of the light degrees of freedom.

Extending this symmetry from charm to bottom, known as heavy-quark flavor symmetry (HQFS), suggests the existence of bottom-strange partners of the $D_{s0}(2317)$ and $D_{s1}(2460)$, namely
the $B_{s0}$ and $B_{s1}$ states, which are expected to exhibit analogous isospin-violating decay modes into $\pi B_s$ and $\pi B_s^{*}$, respectively.  If the $D_{s0}(2317)$ and $D_{s1}(2460)$
are interpreted as $KD$ and $KD^{*}$ molecular states, respectively, symmetry arguments naturally predict that the corresponding bottom--strange partners, $B_{s0}$ and $B_{s1}$, should appear as
$K\bar{B}$ and $K\bar{B}^{*}$ molecular configurations~\cite{Guo:2006fu,Guo:2006rp,Wang:2019ehs,Cheng:2014bca,Colangelo:2012xi,Kolomeitsev:2003ac,Yang:2022vdb,Lang:2015hza}.  However, dedicated
experimental searches have been performed in the theoretically expected isospin-violating decay channels $\pi B_{s}^{(*)}$, yet no clear experimental evidence for such states has been observed so far~\cite{LHCb:2016dxl,CMS:2017hfy,ATLAS:2018udc,CDF:2017dwr}.  This discrepancy between symmetry-based expectations and experimental results raises important questions about the applicability and
limitations of heavy-quark symmetry across different flavor sectors.

It is widely recognized that heavy-quark symmetry is explicitly broken across different flavor sectors due to the finite heavy-quark mass. However, its precise mechanism and quantitative size are
still not fully understood.  This symmetry breaking has been systematically studied within the framework of heavy-quark effective theory (HQET) through an expansion in powers of $1/m_Q$~\cite{Neubert:1993mb, Falk:1992cx, Luke:1992cs}.  It is found that the symmetry-breaking effects originate from the kinetic-energy and chromomagnetic terms, which generate spin-dependent interactions and induce mass splittings
among heavy hadrons.  Beyond HQET, phenomenological and effective field theory approaches have been extensively employed to estimate these corrections in specific systems~\cite{Meng:2022ozq,Endo:2026afe,Becirevic:2004uv,Cheng:2017oqh,Albertus:2005vd}.  In particular, lattice QCD calculations of leptonic $B$, $D$, $B^*$, $D^*$ decays and semileptonic $B \to D^{(*)}$
transitions show that heavy-quark flavor symmetry breaking is typically of order $\sim 10\%$~\cite{Albertus:2005vd}.

Although heavy-quark flavor symmetry breaking effects have been widely studied, their application to the prediction of $K\bar{B}$ and $K\bar{B}^*$ molecular states, which are often associated with
the experimentally unobserved $B_{s0}$ and $B_{s1}$, remains challenging. In particular, such analyses typically involve undetermined model parameters, including the cutoff scale $\Lambda$, the unknown
coupling $g'$, and the parameter $\lambda_1^S$ entering the $1/m_Q$ corrections; see, e.g., Ref.~\cite{Cheng:2017oqh}. These ingredients introduce uncontrolled systematic uncertainties and can induce mass
shifts of order $\mathcal{O}(100\,\mathrm{MeV})$.  A reliable determination of these states therefore requires additional experimental inputs to constrain the relevant parameters.  A possible strategy is
to systematically explore the $D_s$ and $B_s$ spectra for additional states that can be interpreted as $K^{(*)}D^{(*)}$ and $K^{(*)}\bar{B}^{(*)}$ molecular configurations, forming heavy-quark flavor
symmetry (HQSS) partner states. These states can then serve as inputs to constrain the relevant model parameters in a consistent molecular picture.  First, we investigate whether there exist additional
$D_s$-family molecular states beyond the already observed $D_{s0}(2317)$ and $D_{s1}(2460)$.

At present, the PDG lists the ground-state $D_s$, the first excited vector state $D_s^*$, the previously discussed $D_{s0}(2317)$ and $D_{s1}(2460)$, and there are also seven additional excited $D_s$ states~\cite{ParticleDataGroup:2024cfk}.  They are given by $D_{s1}(2536)$, $D_{s2}(2573)$, $D_{s1}^*(2700)$, $D_{s1}(2860)$, $D_{s3}(2860)$, and $D_{sJ}(3040)$.  The states $D_{s1}(2536)$ and $D_{s2}(2573)$
are well established as narrow $P$-wave $c\bar{s}$ excitations with well-determined quantum numbers~\cite{ParticleDataGroup:2024cfk}.  $D_{s1}^*(2700)$ is commonly interpreted as a conventional $c\bar{s}$
meson, more specifically as a mixture of the $2\,^3S_1(c\bar{s})$ and $1\,^3D_1(c\bar{s})$ configurations~\cite{Li:2010vx,Close:2006gr,Godfrey:2013aaa}. There is currently no clear evidence supporting a
tetraquark interpretation, as such assignments typically lead to a sizable mass deviation of order $\mathcal{O}(1\,\mathrm{GeV})$~\cite{Wang:2007nfa}. But, one calculation indicate that
this state may contain a substantial $D K^{*}$ molecular component, which can be as large as approximately $51.3\%$~\cite{Hao:2022vwt}.  The $D_{s1}(2860)$ and $D_{s3}(2860)$ form a spin doublet
consistent with higher orbital excitations~\cite{Wang:2014jua,Song:2014mha,LHCb:2014ott,Chen:2009zt}.  Nevertheless, the $D_{s1}(2860)$ has also been interpreted as a $K^{*}D^{*}$~\cite{Hao:2022vwt} or $D_1(2420)K$~\cite{Guo:2017jvc,Wang:2025jcq,Guo:2011dd} hadronic molecular state, while the $D_{s3}(2860)$ may likewise contain a sizable $K^{*}D^{*}$ molecular component~\cite{Hao:2022vwt}.  The broad structure
$D_{sJ}(3040)$ remains controversial, as it can be interpreted either as a conventional quark-antiquark state~\cite{Chen:2009zt} or as a hadronic molecular configuration~\cite{Hao:2022vwt,Guo:2011dd}.

From the above discussion, it is evident that $D_{s1}^*(2700)$, $D_{s1}(2860)$, and $D_{s3}(2860)$ may contain sizable $K^{*}D$ and $K^{*}D^{*}$ molecular components, respectively, rather than being purely
molecular states. In this work, we investigate a limiting scenario in which these states are assumed to be dominated by pure $K^{*}D$ and $K^{*}D^{*}$ molecular configurations, in order to explore under what
conditions the $D_{s1}^*(2700)$, $D_{s1}(2860)$, and $D_{s3}(2860)$ states can be interpreted as molecular states. Together with the fact that $D_{s0}(2317)$ and $D_{s1}(2460)$ can be interpreted as pure $KD$
and $KD^{*}$ molecular states, respectively, this provides an extreme benchmark for constraining heavy-quark flavor symmetry breaking effects, provided that the existence of $K^{*}\bar{B}$ and $K^{*}\bar{B}^{*}$
molecular states can be confirmed in the future. Recent studies suggest that the state $B_{sJ}(6114)$ may already serve as a candidate for such $K^{*}\bar{B}$ and $K^{*}\bar{B}^{*}$ molecular
configurations~\cite{Sanchez-Illana:2026cyv}.

This paper is organized as follows. In Sec.~\ref{Sec: formulism},  we will present the theoretical  formalism.  In Sec.~\ref{Sec: results}, the numerical result will be given, followed by discussions and
conclusions in the last section.

\section{FORMALISM AND INGREDIENTS}\label{Sec: formulism}
In the present work, we investigate whether the $DK^*$ and $D^*K^*$ interactions can generate molecular states associated with the $D_{s1}^{*}(2700)$ and $D_{s1/s3}^{*}(2860)$, respectively.
Importantly, the coupled-channel effects between these two systems need to be taken into account, as $D$ and $D^{*}$ form a heavy-quark spin symmetry (HQSS) doublet.
To examine the formation of such bound systems, we solve the nonrelativistic Schr\"odinger equation,
\begin{equation}
\left[-\frac{1}{2\mu}\left(\nabla_r^2-\frac{L(L+1)}{r^2}\right)+V(r)\right]\psi(\vec r)
=E\,\psi(\vec r),
\label{eq:schrodinger}
\end{equation}
where the reduced mass is defined as  $\mu = (m_{K^{*}} m_{D{(*)}})/(m_{K^{*}} + m_{D^{(*)}})$.  The radial differential operator takes the form $
\nabla_r^2 = \frac{1}{r^2} \frac{\partial}{\partial r} \left( r^2 \frac{\partial}{\partial r} \right)$.  In this framework, $L$ represents the relative orbital angular momentum between the
two constituents, and the case $L=0$ corresponds to an S-wave configuration.  The wave function $\psi(\vec{r})$ represents the radial component of the $D^{(*)}K^{*}$ molecule, while $E$
corresponds to the binding energy. Accordingly, the mass of the resulting bound state can be written as $m = m_{K^{*}} + m_{D^{(*)}} - E$.

\begin{figure}[http]
\begin{center}
\includegraphics[ width=6cm]{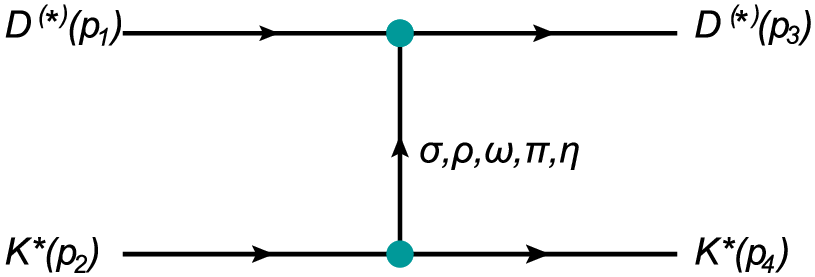}
\caption{Feynman diagram for the process $D^{(*)}K^* \to D^{(*)}K^*$, the four-momenta of the particles are denoted as $p_1$, $p_2$, $p_3$, and $p_4$, respectively.}\label{fig:dsks}
\end{center}
\end{figure}

The binding energy $E$ is determined by the explicit form of the two-body interaction potential $V(r)$ for the $D^{(*)}K^*$ system. In the present work, the $D^{(*)}K^*$ interaction potential
is constructed within the one-boson-exchange (OBE) framework.  At the hadronic level, the interaction kernel for the $D^{(*)}K^* \to D^{(*)}K^*$ process is obtained from
the tree-level diagrams shown in Fig.~\ref{fig:dsks}.   Specifically, for the elastic $DK^* \to DK^*$ channel, the exchanges of the $\sigma$, $\rho$, and $\omega$ mesons are included.  For the
$D^*K^* \to D^*K^*$ channel, the exchanges of the $\pi$, $\eta$, $\rho$, and $\omega$ mesons are taken into account.

To compute the potential $V(r)$ corresponding to Fig.~\ref{fig:dsks}, the following effective Lagrangians, which describe the interaction vertices, are required~\cite{Wang:2024ukc}:
\begin{align}
\mathcal{L}_{D^{(*)}D^{(*)}\sigma}
    &= -2g_s D_b^\dagger D_b \sigma + 2g_s D_b^{*\mu}{D_b^{*\dagger}}_\mu \sigma,
    \label{eq:lD} \\[2mm]
\mathcal{L}_{D^{(*)}D^{(*)}P}
    &= \frac{2g}{f_\pi} (D_b D_a^{*\lambda\dagger}+ D_b^{*\lambda} D_a^\dagger) \partial_\lambda P_{ba}
    \nonumber \\
    &\quad + i \frac{2g}{f_\pi} v^\alpha \epsilon_{\alpha\mu\nu\lambda} D_b^{*\mu} D_a^{*\lambda\dagger} \partial^\nu P_{ba},
    \label{eq:lp} \\[2mm]
    \mathcal{L}_{D^{(*)}D^{(*)}V}
    &= \sqrt{2} \beta g_V D_b D_a^\dagger v \cdot V_{ba}
       - 2\sqrt{2} \lambda g_V \epsilon_{\lambda\mu\alpha\beta} v^\lambda
   + (D_b D_a^{*\mu\dagger}   \nonumber \\
    &\quad+ D_b^{*\mu} D_a^\dagger) (\partial^\alpha V^\beta_{ba})
       - \sqrt{2} \beta g_V D_b^* \cdot D_a^{*\dagger} v \cdot V_{ba}
    \nonumber \\
   &\quad - i 2\sqrt{2} \lambda g_V D_b^{*\mu} D_a^{*\nu\dagger} (\partial_\mu V_\nu - \partial_\nu V_\mu)_{ba},
    \label{eq:lv}\\
    \mathcal{L}_{K^{(*)}K^{(*)}\sigma}
    &= -2g'_s K_b^\dagger K_b \sigma + 2g'_s K_b^{*\mu} {K_b^{*\dagger}}_\mu \sigma,\label{eq:lk}
    \end{align}
    \begin{align}
    \mathcal{L}_{K^{(*)}K^{(*)}P} &= \frac{2g'}{f_\pi} (K_b K_a^{*\lambda\dagger} + K_b^{*\lambda} K_a^\dagger) \partial_\lambda P_{ba}\nonumber \\
    &\quad + i \frac{2g'}{f_\pi} v'^\alpha \epsilon_{\alpha\mu\nu\lambda} K_b^{*\mu} K_a^{*\lambda\dagger} \partial^\nu P_{ba},
    \label{eq:lkp}  \\[2mm]
    \mathcal{L}_{K^{(*)}K^{(*)}V}
    &= \sqrt{2} \beta' g'_V K_b K_a^\dagger v' \cdot V_{ba}
       - 2\sqrt{2} \lambda' g'_{V} \epsilon_{\lambda\mu\alpha\beta} v'^\lambda
  + (K_b K_a^{*\mu\dagger}   \nonumber\\
    &\quad + K_b^{*\mu} K_a^\dagger) (\partial^\alpha V^\beta_{ba})
       - \sqrt{2} \beta' g'_V K_b^* \cdot K_a^{*\dagger} v' \cdot V_{ba}
    \nonumber \\
    &\quad - i 2\sqrt{2} \lambda' g'_V K_b^{*\mu} K_a^{*\nu\dagger} (\partial_\mu V_\nu - \partial_\nu V_\mu)_{ba},
    \label{eq:lkv}
\end{align}
where $\epsilon^{\alpha\mu\nu\lambda}$ denotes the Levi--Civita tensor with $\epsilon^{0123}=1$. The fields $V^{\mu}$ and $P$ represent the SU(3) vector-meson and pseudoscalar-meson
matrices, respectively, given by
\begin{align}
V_{\mu}&=
\left(
  \begin{array}{ccc}
    \frac{1}{\sqrt{2}}(\rho^{0}+\omega) & \rho^{+} & K^{*+} \\
    \rho^{-} & \frac{1}{\sqrt{2}}(-\rho^{0}+\omega) & K^{*0} \\
    K^{*-} & \bar{K}^{*0} & \phi
  \end{array}
\right)_{\mu},\\
P&=
\left(
  \begin{array}{ccc}
    \frac{1}{\sqrt{2}}\pi^{0}+\frac{1}{\sqrt{6}}\eta & \pi^{+} & K^{+} \\
    \pi^{-} & -\frac{1}{\sqrt{2}}\pi^{0}+\frac{1}{\sqrt{6}}\eta & K^{0} \\
    K^{-} & \bar{K}^{0} & -\frac{2}{\sqrt{6}}\eta
  \end{array}
\right).
\end{align}
The coupling constants $g = 0.59 \pm 0.07$ and $g' = 1.12 \pm 0.01$ are obtained from the experimental decay widths of $D^* \to D\pi$ and $K^* \to K\pi$, respectively, by combining the
effective Lagrangians given in Eqs.~\ref{eq:lp} and \ref{eq:lkp}, where $f_\pi = 132~\mathrm{MeV}$ is the pion decay constant appearing in the Lagrangian.  The values $g_V = g'_V = m_\rho / f_\pi = 5.8$
are taken from Refs.~\cite{Isola:2003,Bando:1988}. The parameter $\beta = 0.9$ is determined within the vector meson dominance model, while $\lambda = 0.56~\mathrm{GeV}^{-1}$ is fixed by
matching the form factor obtained from light-cone sum rule calculations to lattice QCD results~\cite{Isola:2003,Bando:1988}. In the strange sector, $\beta' = 0.835$ is obtained from hidden-gauge
symmetry~\cite{Molina:2010}, which is very close to the corresponding value of $\beta$ in the charmed meson sector. Consequently, we adopt $\lambda' = \lambda$ in this work, and the impact of
this assumption on the numerical results is also discussed. Finally, for $\sigma$-meson exchange, the coupling constant $g_s = g'_s = 0.76$ is taken from Ref.~\cite{Liu:2008}.

Because hadrons are not pointlike particles, we need to include the form factors in evaluating the scattering amplitudes of the $\bar{D}^{(*)} K^* \to \bar{D}^{(*)} K^*$ reaction. For the
$t$-channel mesons exchange, we would like to apply a widely used pole form factor, which is
\[\mathcal{F}_i = \frac{\Lambda_i^2 - m_i^2}{\Lambda_i^2 - q_i^2}, \quad i = \pi, \rho, \omega, \sigma, \eta,\] where $m_i$ and $ q_i $ are the mass
and four-momentum of the $i$-th exchanged meson, respectively. The cutoff parameter $\Lambda_i$ is taken as $ \Lambda_i = m_i + \alpha \Lambda_{\text{QCD}} $,  with $ \Lambda_{\text{QCD}} = 220 \, \text{MeV} $,
where \( \alpha \) is treated as a free parameter and will be discussed later.

With the ingredients introduced above, we are now in a position to write down the general expressions for the scattering amplitudes corresponding to the Feynman diagrams shown in Fig.~\ref{fig:dsks}.
The explicit expressions are given as follows:
\begin{align}
\mathcal{M}^{DK^*\to DK^*}_\sigma &= -4g_s g_s' \frac{i}{q^2 - m_\sigma^2} (\epsilon_2 \cdot \epsilon_4^\dagger)\mathcal{F}_{\sigma}^2 \, C_\sigma, \label{eq10}
\end{align}
\begin{align}
\mathcal{M}^{DK^*\to DK^*}_{\rho/\omega} &= -2\beta\beta' g_V g_V' v_\mu \frac{i(-g^{\mu\nu} + q^\mu q^\nu / m_{\rho/\omega}^2)}{q^2 - m_{\rho/\omega}^2} \nonumber\\&(\epsilon_2 \cdot \epsilon_4^\dagger) v'_\nu\mathcal{F}_{\rho/\omega}^2 \, C_{\rho/\omega},\label{eq11}\\
\mathcal{M}^{D^*K^*\to D^*K^*}_\sigma
    &= 4g_s g_s' (\epsilon_1\cdot\epsilon_3^\dagger) \frac{i}{q^2 - m_\sigma^2} (\epsilon_2\cdot\epsilon_4^\dagger) \mathcal{F}_{\sigma}^2\,C_\sigma,
    \label{eq:dksigma} \\[2mm]
\mathcal{M}^{D^*K^*\to D^*K^*}_{\pi/\eta}
    &= \frac{-4gg'}{f_\pi^2} \,
       \epsilon_{ijk} \epsilon_1^i \epsilon_3^{k\dagger} q^j \,
       \frac{i}{q^2 - m_{\pi/\eta}^2}
    \nonumber \\
    &\qquad \times \epsilon_{ijk} \epsilon_2^i \epsilon_4^{k\dagger} q^j \,
       \mathcal{F}_{\pi/\eta}^2\,C_{\pi/\eta},
    \label{eq:dkpi} \\[2mm]
\mathcal{M}^{D^*K^*\to D^*K^*}_{\rho/\omega}
    &= \frac{i\bigl(-g^{\mu\nu} + q^\mu q^\nu / m_{\rho/\omega}^2\bigr)}
            {q^2 - m_{\rho/\omega}^2}
       \Biggl[
       2\beta\beta' g_V g_V' (\epsilon_1\cdot\epsilon_3^\dagger) v_\mu
          \nonumber \\
    &\qquad(\epsilon_2\cdot\epsilon_4^\dagger) v'_\nu
     + 8\lambda\lambda' g_V g_V' \,
          \epsilon_1^\alpha \epsilon_3^{\beta\dagger}
          \bigl(q_{\beta} g_{\mu\alpha} - q_{\alpha} g_{\mu\beta}\bigr)
    \nonumber \\
    &\qquad \times \epsilon_2^\theta \epsilon_4^{\lambda\dagger}
          \bigl(q_{\theta} g_{\nu\lambda} - q_{\lambda} g_{\nu\theta}\bigr)
       \Biggr]\mathcal{F}_{\rho/\omega}^2 \times C_{\rho/\omega},
    \label{eq:dkrho}\\
\mathcal{M}^{DK^*\to D^*K^*}_{\pi/\eta}
    &= \frac{4i gg'}{f_\pi^2}\,
       \epsilon_{3\lambda}^\dagger q^\lambda \,
       \frac{i}{q^2 - m_{\pi/\eta}^2}\,
       \epsilon_{ijk} \epsilon_2^i \epsilon_4^{k\dagger} q^j\,\mathcal{F}_{\pi/\eta}^2
       C_{\pi/\eta},
       \label{eq:dkdkstar_pi} \\[2mm]
\mathcal{M}^{DK^*\to D^*K^*}_{\rho/\omega}
    &= 8i\lambda\lambda' g_V g_V' \,
       \epsilon_{ijk} \epsilon_3^{i\dagger} q^j g^{\mu k}
       \frac{i\bigl(-g^{\mu\nu} + q^\mu q^\nu / m_{\rho/\omega}^2\bigr)}
            {q^2 - m_{\rho/\omega}^2}
       \nonumber \\
    &\qquad \times \epsilon_2^\alpha \epsilon_4^{\beta\dagger}
       \bigl(q_\alpha g_{\nu\beta} - q_\beta g_{\nu\alpha}\bigr)\,\mathcal{F}_{\rho/\omega}^2\,
       C_{\rho/\omega},
       \label{eq:dkdkstar_rho}
\end{align}
where $q = p_3 - p_1 = p_2 - p_4$, and $\epsilon^\mu$ denotes the meson polarization vector, for which we adopt the explicit basis
$\epsilon^\mu(s=0) = (0,0,0,-1)$ and $\epsilon^\mu(s=\pm) = \frac{1}{\sqrt{2}}(0,\pm 1,i,0)$ in our numerical evaluation.  The coefficients
are given by $C_\sigma = 1$, $C_\pi = -\frac{3}{2}$, $C_\eta = \frac{1}{6}$, $C_\rho = -\frac{3}{2}$, and $C_\omega = \frac{1}{2}$, which are
obtained from the following isospin relations.
\begin{align}
|D_{s1}^{*+}(2860)\rangle &= \frac{1}{\sqrt{2}}\left(|D^{*+}K^{*0}\rangle - |D^{*0}K^{*+}\rangle\right), \nonumber \\
|D_{s1}^{*+}(2700)\rangle &= \frac{1}{\sqrt{2}}\left(|D^{+}K^{*0}\rangle - |D^{0}K^{*+}\rangle\right).
\end{align}

Within the Breit approximation, the momentum-space interaction potential is readily obtained. Its Fourier transform yields the coordinate-space potential
$V(\vec{r})$ entering Eq.~\ref{eq:schrodinger}. The Fourier-transform relations employed in this work are summarized below.
\begin{align}
&{\cal F}\!\left\{\frac{1}{\vec{q}^{\,2}+m^{2}}
    \left(\frac{\Lambda^{2}-m^{2}}{\Lambda^{2}+\vec{q}^{\,2}}\right)^2
\right\}= Y_{1}(\Lambda,m,r),\nonumber\\[2mm]
&{\cal F}\!\left\{
    \frac{\vec{q}^{\,2}}{\vec{q}^{\,2}+m^{2}}
    \left(\frac{\Lambda^{2}-m^{2}}{\Lambda^{2}+\vec{q}^{\,2}}\right)^{2}
\right\}
= -\nabla_{r}^{2} Y_{1}(\Lambda,m,r), \nonumber\\[2mm]
&{\cal F}\!\left\{
    \frac{(\vec{A}\cdot\vec{q})(\vec{B}\cdot\vec{q})}
         {\vec{q}^{\,2}+m^{2}}
    \left(\frac{\Lambda^{2}-m^{2}}{\Lambda^{2}+\vec{q}^{\,2}}\right)^{2}
\right\}
= \frac{1}{3}(\vec{A}\!\cdot\!\vec{B})
  \big(-\nabla_{r}^{2}Y_{1}(\Lambda,m,r)\big) \nonumber\\
&\hspace{3cm}
  + \frac{1}{3} S(\hat{r},\vec{A},\vec{B})
    \left(
      - r\frac{\partial}{\partial r}
        \frac{1}{r}\frac{\partial}{\partial r}
        Y_{1}(\Lambda,m,r)
    \right). \nonumber
\end{align}
Here ${\cal F}$ denotes the Fourier transformation, and
$\nabla_r^{2} = \dfrac{1}{r^{2}} \dfrac{\partial}{\partial r}
\!\left( r^{2} \dfrac{\partial}{\partial r} \right)$
is the radial Laplacian, while
$S(\hat{r},\vec{x},\vec{y}) = 3(\hat{r}\!\cdot\!\vec{x})(\hat{r}\!\cdot\!\vec{y}) - \vec{x}\!\cdot\!\vec{y}$
denotes the tensor operator.  The function $Y_1(\Lambda,m,r)$ is defined as
\begin{align}
&Y_1(\Lambda,m,r) = \frac{1}{4\pi r} ( e^{-m r} - e^{-\Lambda r} )- \frac{\Lambda^2 - m^2}{8 \pi \Lambda} e^{-\Lambda r}.
\end{align}
The differential operators $\partial/\partial r$ and $\nabla_r^2$ act on the radial wave function $\psi(\vec{r})$, which is expanded using the Gaussian expansion method~\cite{Hiyama:2003cu}.
The operators, such as $S(\hat{r},\vec{A},\vec{B})$ and $\vec{\epsilon}_2 \cdot \vec{\epsilon}_4^{\dagger}$, have been evaluated in detail, and the corresponding results
are collected in Tables~\ref{tab:tableqw-1}--\ref{tab:tableqw-2}.  It should be noted that, in addition to the $S$-wave case, higher partial waves are also considered, with the total
spin-parity quantum numbers ranging from $J^P = 0^{+}$ to $J^P = 3^{+}$, as summarized in Table~\ref{tab:spins}.

\begin{table}[htbp]
\caption{Possible quantum numbers for $D^{(*)}K^*$ systems involved in our calculation.
The first column contains the spin-parity quantum numbers corresponding to the channels.
$A \sim B$ stands for the mixing effect between $A$ and $B$.}\label{tab:spins}
\centering
\setlength{\tabcolsep}{16pt}
\begin{tabular}{c c c}
\hline
$J^P$ & $D^*{K}^*$ states & $D{K}^*$ states \\
\hline
$0^+$ & $|^1 S_0 \sim {}^5 D_0\rangle$ & --- \\
$0^-$ & $|^3 P_0\rangle$ & $|^3 P_0\rangle$ \\
$1^+$ & $|^3 S_1 \sim {}^3 D_1 \sim {}^5 D_1\rangle$ & $|^3 S_1 \sim {}^3 D_1\rangle$ \\
$1^-$ & $|^1 P_1 \sim {}^3 P_1 \sim {}^5 P_1 \sim {}^5 F_1\rangle$ & $|^3 P_1\rangle$ \\
$2^+$ & $|^1 D_2 \sim {}^3 D_2 \sim {}^5 S_2 \sim {}^5 D_2\rangle$ & $|^3 D_2\rangle$ \\
$2^-$ & $|^3 P_2 \sim {}^3 F_2 \sim {}^5 P_2 \sim {}^5 F_2\rangle$ & $|^3 P_2 \sim {}^3 F_2\rangle$ \\
$3^+$ & $|^3 D_3 \sim {}^5 D_3\rangle$ & $|^3 D_3\rangle$ \\
$3^-$ & $|^1 F_3 \sim {}^3 F_3 \sim {}^5 P_3 \sim {}^5 F_3\rangle$ & $|^3 F_3\rangle$ \\
\hline
\end{tabular}
\end{table}

\begin{table*}[t!]
	\centering
	\caption{The matrix elements of two-body interaction operators for $D^{(*)}K^*$ systems.}\label{tab:tableqw-1}
     \setlength{\tabcolsep}{1.5pt}
	\begin{tabular}{cccccccc}
		\hline\hline
		$D^*K^*\to{}D^*K^*$    &$0^{+}$                    & $0^{-}$                               &$1^{-}$                  &$1^{+}$         \\
${\cal{Z}}$		&($^{1}S_{J}$,$^{5}D_{J}$)& (${}^{3}P_{J}$) &($^1P_{J},{}^{3}P_{J}, {}^{5}P_{J},{}^5F_{J}$)
&($^3S_{J},{}^{3}D_{J}, {}^{5}D_{J}$)  \\
	$\vec{\epsilon}_1\cdot{}\vec{\epsilon}_2S(\hat{r},\vec{\epsilon}_3^{\dagger},\vec{\epsilon}_{4}^{\dagger})$&
		$\left(
		\begin{array}{cc}
			0 & -\frac{\sqrt{2}}{5} \\
-\frac{\sqrt{2}}{5} & \frac{2}{35}\\
		\end{array}
		\right)$  &
		$\left(
		\begin{array}{c}
			-\frac{2}{15}  \\
		\end{array}
		\right)$ &
		$\left(
		\begin{array}{cccc}
		0 & 0 & \frac{4}{5\sqrt{5}} & -\frac{9\sqrt{\frac{6}{5}}}{35} \\
0 & 0 & 0 & 0 \\
\frac{4}{5\sqrt{5}} & 0 & \frac{4}{25} & -\frac{9\sqrt{6}}{175} \\
-\frac{9\sqrt{\frac{6}{5}}}{35} & 0 & -\frac{9\sqrt{6}}{175} & \frac{12}{175}
		\end{array}
		\right)$&
		$\left(
		\begin{array}{ccc}
			0 & \frac{\sqrt{2}}{5} & 0 \\
\frac{\sqrt{2}}{5} & -\frac{4}{35} & 0 \\
0 & 0 & 0
		\end{array}
		\right)$
		\\	
$\vec{\epsilon}_1\cdot{}\vec{\epsilon}_{3}^{\dagger}S(\hat{r},\vec{\epsilon}_2,\vec{\epsilon}_{4}^{\dagger})$&
		$\left(
		\begin{array}{cc}
			0 & -\frac{\sqrt{2}}{5} \\
-\frac{\sqrt{2}}{5} & \frac{1}{5}
		\end{array}
		\right)$
		& $\left(
			\begin{array}{c}
			   -\frac{1}{5}\\
			\end{array}
			\right)$
		& $\left(
			\begin{array}{cccc}
			0 & 0 & \frac{4}{5\sqrt{5}} & -\frac{9\sqrt{\frac{6}{5}}}{35} \\
0 & -\frac{1}{10} & -\frac{3\sqrt{\frac{3}{5}}}{10} & -\frac{18\sqrt{\frac{2}{5}}}{35} \\
\frac{4}{5\sqrt{5}} & -\frac{3\sqrt{\frac{3}{5}}}{10} & \frac{1}{10} & -\frac{3\sqrt{6}}{35} \\
-\frac{9\sqrt{\frac{6}{5}}}{35} & -\frac{18\sqrt{\frac{2}{5}}}{35} & -\frac{3\sqrt{6}}{35} & \frac{4}{35}
			\end{array}
			\right)$
		&$\left(\begin{array}{ccc}
				0 & \frac{\sqrt{2}}{5} & 0 \\
\frac{\sqrt{2}}{5} & -\frac{1}{14} & -\frac{3\sqrt{3}}{70} \\
0 & -\frac{3\sqrt{3}}{70} & \frac{17}{70}\\
			\end{array}\right) $\\
$\vec{\epsilon}_2\cdot{}\vec{\epsilon}_{4}^{\dagger}S(\hat{r},\vec{\epsilon}_1,\vec{\epsilon}_{3}^{\dagger})$&

		$\left(
		\begin{array}{cc}
			0 & -\frac{\sqrt{2}}{5} \\
-\frac{\sqrt{2}}{5} & \frac{1}{5}
		\end{array}
		\right)$
		& $\left(
			\begin{array}{c}
			   -\frac{1}{5}
			\end{array}
			\right)$
		& $\left(
			\begin{array}{cccc}
			0 & 0 & \frac{4}{5\sqrt{5}} & -\frac{9\sqrt{\frac{6}{5}}}{35} \\
0 & -\frac{1}{10} & \frac{3\sqrt{\frac{3}{5}}}{10} & \frac{18\sqrt{\frac{2}{5}}}{35} \\
\frac{4}{5\sqrt{5}} & \frac{3\sqrt{\frac{3}{5}}}{10} & \frac{1}{10} & -\frac{3\sqrt{6}}{35} \\
-\frac{9\sqrt{\frac{6}{5}}}{35} & \frac{18\sqrt{\frac{2}{5}}}{35} & -\frac{3\sqrt{6}}{35} & \frac{4}{35}
			\end{array}
			\right)$
		&$\left(\begin{array}{ccc}
				0 & \frac{\sqrt{2}}{5} & 0 \\
\frac{\sqrt{2}}{5} & -\frac{1}{14} & \frac{3\sqrt{3}}{70} \\
0 & \frac{3\sqrt{3}}{70} & \frac{17}{70}
			\end{array}\right) $\\
$\vec{\epsilon}_1\cdot{}\vec{\epsilon}_{4}^{\dagger}S(\hat{r},\vec{\epsilon}_2,\vec{\epsilon}_{3}^{\dagger})$&
          $\left(
		\begin{array}{cc}
			0 & -\frac{2\sqrt{2}}{15} \\
-\frac{2\sqrt{2}}{15} & \frac{4}{21}
		\end{array}
		\right)$
            & $\left(
			\begin{array}{c}
			    0
			\end{array}
			\right)$
           & $\left(
			\begin{array}{cccc}
			\frac{4}{15} & 0 & \frac{8}{15\sqrt{5}} & -\frac{6\sqrt{\frac{6}{5}}}{35} \\
0 & 0 & 0 & 0 \\
\frac{8}{15\sqrt{5}} & 0 & \frac{16}{75} & -\frac{12\sqrt{6}}{175} \\
-\frac{6\sqrt{\frac{6}{5}}}{35} & 0 & -\frac{12\sqrt{6}}{175} & \frac{16}{175}
			\end{array}
			\right)$
        &$\left(\begin{array}{ccc}
				0 & 0 & 0 \\
0 & 0 & 0 \\
0 & 0 & \frac{8}{35}
			\end{array}\right) $\\
$(\vec{\epsilon}_1\cdot{}\vec{\epsilon}_{2})(\vec{\epsilon}_{3}^{\dagger}\cdot{}\vec{\epsilon}_{4}^{\dagger})$
&$\left(
		\begin{array}{cc}
			1 & 0 \\
0 & \frac{1}{5}
		\end{array}
		\right)$
            & $\left(
			\begin{array}{c}
			    \frac{1}{3}
			\end{array}
			\right)$
           & $\left(
			\begin{array}{cccc}
			1 & 0 & 0 & 0 \\
0 & 0 & 0 & 0 \\
0 & 0 & \frac{2}{5} & 0 \\
0 & 0 & 0 & \frac{9}{35}
			\end{array}
			\right)$
        &$\left(\begin{array}{ccc}
				1 & 0 & 0 \\
0 & \frac{2}{5} & 0 \\
0 & 0 & 0
			\end{array}\right) $\\
$(\vec{\epsilon}_2\cdot{}\vec{\epsilon}_{3}^{\dagger})(\vec{\epsilon}_{1}\cdot{}\vec{\epsilon}_{4}^{\dagger})$
&$\left(
		\begin{array}{cc}
			\frac{1}{3} & 0 \\
0 & \frac{8}{15}
		\end{array}
		\right)$
            & $\left(
			\begin{array}{c}
			    0
			\end{array}
			\right)$
           & $\left(
			\begin{array}{cccc}
			\frac{1}{3} & 0 & \frac{2}{3\sqrt{5}} & 0 \\
0 & 0 & 0 & 0 \\
\frac{2}{3\sqrt{5}} & 0 & \frac{4}{15} & 0 \\
0 & 0 & 0 & \frac{16}{35}
			\end{array}
			\right)$
        &$\left(\begin{array}{ccc}
				0 & 0 & 0 \\
0 & 0 & 0 \\
0 & 0 & \frac{4}{5}
			\end{array}\right) $\\
		
   $(\vec{\epsilon}_2\cdot{}\vec{\epsilon}_{4}^{\dagger})(\vec{\epsilon}_{1}\cdot{}\vec{\epsilon}_{3}^{\dagger})$
&$\left(
		\begin{array}{cc}
			1 & 0 \\
0 & 1
		\end{array}
		\right)$
            & $\left(
			\begin{array}{c}
			    1
			\end{array}
			\right)$
           & $\left(
			\begin{array}{cccc}
			1 & 0 & 0 & 0 \\
0 & 1 & 0 & 0 \\
0 & 0 & 1 & 0 \\
0 & 0 & 0 & 1
			\end{array}
			\right)$
        &$\left(\begin{array}{ccc}
				1 & 0 & 0 \\
0 & 1 & 0 \\
0 & 0 & 1
			\end{array}\right) $\\

    $S(\hat{r},\vec{\epsilon}_1\times\vec{\epsilon}_3^{\dagger},\vec{\epsilon}_2\times\vec{\epsilon}_{4}^{\dagger})$&
		$\left(
		\begin{array}{cc}
			0 & \frac{2\sqrt{2}}{15} \\
\frac{2\sqrt{2}}{15} & \frac{4}{15}
		\end{array}
		\right)$  &
		$\left(
		\begin{array}{c}
			\frac{4}{15}
		\end{array}
		\right)$ &
		$\left(
		\begin{array}{cccc}
		\frac{8}{15} & 0 & -\frac{8}{15\sqrt{5}} & \frac{6\sqrt{\frac{6}{5}}}{35} \\
0 & 0 & 0 & 0 \\
-\frac{8}{15\sqrt{5}} & 0 & \frac{8}{75} & -\frac{6\sqrt{6}}{175} \\
\frac{6\sqrt{\frac{6}{5}}}{35} & 0 & -\frac{6\sqrt{6}}{175} & \frac{8}{175}
		\end{array}
		\right)$&
		$\left(
		\begin{array}{ccc}
			0 & -\frac{2\sqrt{2}}{5} & 0 \\
-\frac{2\sqrt{2}}{5} & \frac{8}{35} & 0 \\
0 & 0 & \frac{16}{35}
		\end{array}
		\right)$
		\\	

    $(\vec{\epsilon}_1\times\vec{\epsilon}_3^{\dagger})(
    \vec{\epsilon}_2\times\vec{\epsilon}_{4}^{\dagger})$&
		$\left(
		\begin{array}{cc}
			\frac{2}{3} & 0 \\
0 & -\frac{1}{3}
		\end{array}
		\right)$  &
		$\left(
		\begin{array}{c}
			\frac{1}{3}
		\end{array}
		\right)$ &
		$\left(
		\begin{array}{cccc}
		\frac{2}{3} & 0 & -\frac{2}{3\sqrt{5}} & 0 \\
0 & 0 & 0 & 0 \\
-\frac{2}{3\sqrt{5}} & 0 & \frac{2}{15} & 0 \\
0 & 0 & 0 & -\frac{1}{5}
		\end{array}
		\right)$&
		$\left(
		\begin{array}{ccc}
			1 & 0 & 0 \\
0 & \frac{2}{5} & 0 \\
0 & 0 & -\frac{4}{5}
		\end{array}
		\right)$
		\\	\hline
$DK^*\to{}DK^*$  &  &$0^{-}$                    & $1^{+}$                               &$1^{-}$            \\
${\cal{Z}}$&	&$^{3}P_{J}$& ($^{3}S_{J},^{3}D_{J}$) &$^{3}P_J$ \\$\vec{\epsilon}_{2}\cdot{}\vec{\epsilon}_{4}^{\dagger}$
& &$\left(
			\begin{array}{c}
			  1
			\end{array}
			\right)$
&$\left(
			\begin{array}{cc}
			  1&0\\0&1
			\end{array}
			\right)$
&$\left(
			\begin{array}{c}
			  1
			\end{array}
			\right)$\\\hline\hline
\end{tabular}
\end{table*}

\begin{table*}[t!]
	\centering
	\caption{The matrix elements of two-body interaction operators for $D^{(*)}K^*$ systems.}\label{tab:tableqw-2}
     \setlength{\tabcolsep}{1.5pt}
	\begin{tabular}{ccccc}\hline\hline
 $D^*K^*\to{}D^*K^*$    &$2^{+}$                    & $2^{-}$                               &$3{-}$                   &$3^{+}$                   \\
${\cal{Z}}$		&($^{1}D_{J},^{3}D_{J},^{5}S_{J},^{5}D_{J}$)& ($^{3}P_{J},^{3}F_{J},^{5}P_{J},^{5}F_{J}$) &($^{1}F_J, ^{3}F_{J}, ^{5}P_{J}, ^{5}F_{J}$) &($^{3}D_{J},^{5}D_{J}$) \\
$\vec{\epsilon}_1\cdot{}\vec{\epsilon}_{2}^{\dagger}S(\hat{r},\vec{\epsilon}_{3}^{\dagger},\vec{\epsilon}_{4}^{\dagger})$&
$\left(
		\begin{array}{cccc}
			0 & 0 & -\sqrt{\frac{2}{5}} & \frac{4}{7\sqrt{7}} \\
0 & 0 & 0 & 0 \\
-\sqrt{\frac{2}{5}} & 0 & 0 & -\sqrt{\frac{2}{35}} \\
\frac{4}{7\sqrt{7}} & 0 & -\sqrt{\frac{2}{35}} & \frac{4}{49}
		\end{array}
		\right)$  &
		$\left(
		\begin{array}{cccc}
			-\frac{4}{15} & \frac{3\sqrt{6}}{35} & 0 & 0 \\
\frac{3\sqrt{6}}{35} & -\frac{4}{35} & 0 & 0 \\
0 & 0 & 0 & 0 \\
0 & 0 & 0 & 0
		\end{array}
		\right)$ &
		$\left(
		\begin{array}{cccc}
		0 & 0 & -\frac{9\sqrt{\frac{2}{35}}}{5} & \frac{8\sqrt{\frac{2}{15}}}{15} \\
0 & 0 & 0 & 0 \\
-\frac{9\sqrt{\frac{2}{35}}}{5} & 0 & \frac{6}{25} & -\frac{6\sqrt{\frac{3}{7}}}{25} \\
\frac{8\sqrt{\frac{2}{15}}}{15} & 0 & -\frac{6\sqrt{\frac{3}{7}}}{25} & \frac{16}{225}
		\end{array}
		\right)$&
		$\left(
		\begin{array}{cc}
			-\frac{6}{35} & 0 \\
0 & 0
		\end{array}
		\right)$\\
$\vec{\epsilon}_1\cdot{}\vec{\epsilon}_{3}^{\dagger}S(\hat{r},\vec{\epsilon}_2,\vec{\epsilon}_{4}^{\dagger})$&
		$\left(
		\begin{array}{cccc}
			0 & 0 & -\sqrt{\frac{2}{5}} & \frac{4}{7\sqrt{7}} \\
0 & \frac{1}{14} & 0 & \frac{3}{14\sqrt{7}} \\
-\sqrt{\frac{2}{5}} & 0 & 0 & -\sqrt{\frac{2}{35}} \\
\frac{4}{7\sqrt{7}} & \frac{3}{14\sqrt{7}} & -\sqrt{\frac{2}{35}} & \frac{25}{98}
		\end{array}
		\right)$
		& $\left(
			\begin{array}{cccc}
			   -\frac{3}{10} & \frac{4\sqrt{6}}{35} & -\frac{\sqrt{3}}{10} & -\frac{3\sqrt{3}}{35} \\
\frac{4\sqrt{6}}{35} & -\frac{2}{35} & \frac{9\sqrt{2}}{35} & -\frac{3\sqrt{2}}{35} \\
-\frac{\sqrt{3}}{10} & \frac{9\sqrt{2}}{35} & -\frac{1}{10} & -\frac{3}{35} \\
-\frac{3\sqrt{3}}{35} & -\frac{3\sqrt{2}}{35} & -\frac{3}{35} & \frac{1}{35}
			\end{array}
			\right)$
		& $\left(
			\begin{array}{cccc}
			0 & 0 & -\frac{9\sqrt{\frac{2}{35}}}{5} & \frac{8\sqrt{\frac{2}{15}}}{15} \\
0 & \frac{1}{10} & \frac{9}{5\sqrt{35}} & \frac{\sqrt{\frac{3}{5}}}{10} \\
-\frac{9\sqrt{\frac{2}{35}}}{5} & \frac{9}{5\sqrt{35}} & \frac{1}{5} & -\frac{\sqrt{\frac{3}{7}}}{5} \\
\frac{8\sqrt{\frac{2}{15}}}{15} & \frac{\sqrt{\frac{3}{5}}}{10} & -\frac{\sqrt{\frac{3}{7}}}{5} & \frac{7}{90}
			\end{array}
			\right)$&
		$\left(\begin{array}{cc}
				-\frac{1}{7} & \frac{3\sqrt{2}}{35} \\
\frac{3\sqrt{2}}{35} & \frac{4}{35}
			\end{array}\right) $\\
$\vec{\epsilon}_2\cdot{}\vec{\epsilon}_{4}^{\dagger}S(\hat{r},\vec{\epsilon}_1,\vec{\epsilon}_{3}^{\dagger})$&
	$\left(
		\begin{array}{cccc}
			0 & 0 & -\sqrt{\frac{2}{5}} & \frac{4}{7\sqrt{7}} \\
0 & \frac{1}{14} & 0 & -\frac{3}{14\sqrt{7}} \\
-\sqrt{\frac{2}{5}} & 0 & 0 & -\sqrt{\frac{2}{35}} \\
\frac{4}{7\sqrt{7}} & -\frac{3}{14\sqrt{7}} & -\sqrt{\frac{2}{35}} & \frac{25}{98}
		\end{array}
		\right)$
		& $\left(
			\begin{array}{cccc}
			   -\frac{3}{10} & \frac{4\sqrt{6}}{35} & \frac{\sqrt{3}}{10} & \frac{3\sqrt{3}}{35} \\
\frac{4\sqrt{6}}{35} & -\frac{2}{35} & -\frac{9\sqrt{2}}{35} & \frac{3\sqrt{2}}{35} \\
\frac{\sqrt{3}}{10} & -\frac{9\sqrt{2}}{35} & -\frac{1}{10} & -\frac{3}{35} \\
\frac{3\sqrt{3}}{35} & \frac{3\sqrt{2}}{35} & -\frac{3}{35} & \frac{1}{35}
			\end{array}
			\right)$
		& $\left(
			\begin{array}{cccc}
			0 & 0 & -\frac{9\sqrt{\frac{2}{35}}}{5} & \frac{8\sqrt{\frac{2}{15}}}{15} \\
0 & \frac{1}{10} & -\frac{9}{5\sqrt{35}} & -\frac{\sqrt{\frac{3}{5}}}{10} \\
-\frac{9\sqrt{\frac{2}{35}}}{5} & -\frac{9}{5\sqrt{35}} & \frac{1}{5} & -\frac{\sqrt{\frac{3}{7}}}{5} \\
\frac{8\sqrt{\frac{2}{15}}}{15} & -\frac{\sqrt{\frac{3}{5}}}{10} & -\frac{\sqrt{\frac{3}{7}}}{5} & \frac{7}{90}
			\end{array}
			\right)$
		&$\left(\begin{array}{cc}
				-\frac{1}{7} & -\frac{3\sqrt{2}}{35} \\
-\frac{3\sqrt{2}}{35} & \frac{4}{35}
			\end{array}\right) $\\
$\vec{\epsilon}_1\cdot{}\vec{\epsilon}_{4}^{\dagger}S(\hat{r},\vec{\epsilon}_2,\vec{\epsilon}_{3}^{\dagger})$&
          $\left(
		\begin{array}{cccc}
			\frac{4}{21} & 0 & -\frac{2\sqrt{\frac{2}{5}}}{3} & \frac{8}{21\sqrt{7}} \\
0 & 0 & 0 & 0 \\
-\frac{2\sqrt{\frac{2}{5}}}{3} & 0 & 0 & -\frac{4\sqrt{\frac{2}{35}}}{3} \\
\frac{8}{21\sqrt{7}} & 0 & -\frac{4\sqrt{\frac{2}{35}}}{3} & \frac{40}{147}
		\end{array}
		\right)$
            & $\left(
			\begin{array}{cccc}
			   0 & 0 & 0 & 0 \\
0 & 0 & 0 & 0 \\
0 & 0 & 0 & 0 \\
0 & 0 & 0 & 0
			\end{array}
			\right)$
           & $\left(
			\begin{array}{cccc}
			\frac{8}{45} & 0 & -\frac{6\sqrt{\frac{2}{35}}}{5} & \frac{16\sqrt{\frac{2}{15}}}{45} \\
0 & 0 & 0 & 0 \\
-\frac{6\sqrt{\frac{2}{35}}}{5} & 0 & \frac{8}{25} & -\frac{8\sqrt{\frac{3}{7}}}{25} \\
\frac{16\sqrt{\frac{2}{15}}}{45} & 0 & -\frac{8\sqrt{\frac{3}{7}}}{25} & \frac{64}{675}
			\end{array}
			\right)$
        &$\left(\begin{array}{cc}
				0 & 0 \\
0 & \frac{2}{35}
			\end{array}\right) $\\
$(\vec{\epsilon}_1\cdot{}\vec{\epsilon}_{2})(\vec{\epsilon}_{3}^{\dagger}\cdot{}\vec{\epsilon}_{4}^{\dagger})$
&$\left(
		\begin{array}{cccc}
			1 & 0 & 0 & 0 \\
0 & 0 & 0 & 0 \\
0 & 0 & 1 & 0 \\
0 & 0 & 0 & \frac{2}{7}
		\end{array}
		\right)$
            & $\left(
			\begin{array}{cccc}
			   \frac{2}{3} & 0 & 0 & 0 \\
0 & \frac{3}{7} & 0 & 0 \\
0 & 0 & 0 & 0 \\
0 & 0 & 0 & 0
			\end{array}
			\right)$
           & $\left(
			\begin{array}{cccc}
			1 & 0 & 0 & 0 \\
0 & 0 & 0 & 0 \\
0 & 0 & \frac{3}{5} & 0 \\
0 & 0 & 0 & \frac{4}{15}
			\end{array}
			\right)$
        &$\left(\begin{array}{ccc}
	\frac{3}{5} & 0 \\
0 & 0
			\end{array}\right) $\\
$(\vec{\epsilon}_2\cdot{}\vec{\epsilon}_{3}^{\dagger})(\vec{\epsilon}_{1}\cdot{}\vec{\epsilon}_{4}^{\dagger})$
&$\left(
		\begin{array}{cccc}
			\frac{1}{3} & 0 & 0 & \frac{2}{3\sqrt{7}} \\
0 & 0 & 0 & 0 \\
0 & 0 & \frac{2}{3} & 0 \\
\frac{2}{3\sqrt{7}} & 0 & 0 & \frac{16}{21}
		\end{array}
		\right)$
            & $\left(
			\begin{array}{cccc}
			   0 & 0 & 0 & 0 \\
0 & 0 & 0 & 0 \\
0 & 0 & 0 & 0 \\
0 & 0 & 0 & \frac{5}{7}
			\end{array}
			\right)$
           & $\left(
			\begin{array}{cccc}
			\frac{1}{3} & 0 & 0 & \frac{2\sqrt{\frac{2}{15}}}{3} \\
0 & 0 & 0 & 0 \\
0 & 0 & \frac{2}{5} & 0 \\
\frac{2\sqrt{\frac{2}{15}}}{3} & 0 & 0 & \frac{38}{45}
			\end{array}
			\right)$
        &$\left(\begin{array}{cc}
				0 & 0 \\
0 & \frac{1}{5}
			\end{array}\right) $\\
		
   $(\vec{\epsilon}_2\cdot{}\vec{\epsilon}_{4}^{\dagger})(\vec{\epsilon}_{1}\cdot{}\vec{\epsilon}_{3}^{\dagger})$
&$\left(
		\begin{array}{cccc}
			1 & 0 & 0 & 0 \\
0 & 1 & 0 & 0 \\
0 & 0 & 1 & 0 \\
0 & 0 & 0 & 1
		\end{array}
		\right)$
            & $\left(
			\begin{array}{cccc}
			  1 & 0 & 0 & 0 \\
0 & 1 & 0 & 0 \\
0 & 0 & 1 & 0 \\
0 & 0 & 0 & 1
			\end{array}
			\right)$
           & $\left(
			\begin{array}{cccc}
			1 & 0 & 0 & 0 \\
0 & 1 & 0 & 0 \\
0 & 0 & 1 & 0 \\
0 & 0 & 0 & 1
			\end{array}
			\right)$
        &$\left(\begin{array}{cc}
				1 & 0  \\
0 & 1
			\end{array}\right) $\\
	
    $S(\hat{r},\vec{\epsilon}_1\times\vec{\epsilon}_3^{\dagger},\vec{\epsilon}_2\times\vec{\epsilon}_{4}^{\dagger})$&
		$\left(
		\begin{array}{cccc}
			\frac{8}{21} & 0 & \frac{2\sqrt{\frac{2}{5}}}{3} & -\frac{8}{21\sqrt{7}} \\
0 & 0 & 0 & 0 \\
\frac{2\sqrt{\frac{2}{5}}}{3} & 0 & 0 & -\frac{2\sqrt{\frac{2}{35}}}{3} \\
-\frac{8}{21\sqrt{7}} & 0 & -\frac{2\sqrt{\frac{2}{35}}}{3} & \frac{8}{21}
		\end{array}
		\right)$  &
		$\left(
		\begin{array}{cccc}
			\frac{8}{15} & -\frac{6\sqrt{6}}{35} & 0 & 0 \\
-\frac{6\sqrt{6}}{35} & \frac{8}{35} & 0 & 0 \\
0 & 0 & 0 & 0 \\
0 & 0 & 0 & 0
		\end{array}
		\right)$ &
		$\left(
		\begin{array}{cccc}
		\frac{16}{45} & 0 & \frac{6\sqrt{\frac{2}{35}}}{5} & -\frac{16\sqrt{\frac{2}{15}}}{45} \\
0 & 0 & 0 & 0 \\
\frac{6\sqrt{\frac{2}{35}}}{5} & 0 & \frac{4}{25} & -\frac{4\sqrt{\frac{3}{7}}}{25} \\
-\frac{16\sqrt{\frac{2}{15}}}{45} & 0 & -\frac{4\sqrt{\frac{3}{7}}}{25} & \frac{32}{675}
		\end{array}
		\right)$&
		$\left(
		\begin{array}{cc}
			\frac{12}{35} & 0 \\[2mm]
0 & \frac{4}{35}
		\end{array}
		\right)$
		\\	

    $(\vec{\epsilon}_1\times\vec{\epsilon}_3^{\dagger})(
    \vec{\epsilon}_2\times\vec{\epsilon}_{4}^{\dagger})$&
		$\left(
		\begin{array}{cccc}
			\frac{2}{3} & 0 & 0 & -\frac{2}{3\sqrt{7}} \\
0 & 0 & 0 & 0 \\
0 & 0 & \frac{1}{3} & 0 \\
-\frac{2}{3\sqrt{7}} & 0 & 0 & -\frac{10}{21}
		\end{array}
		\right)$  &
		$\left(
		\begin{array}{cccc}
			\frac{2}{3} & 0 & 0 & 0 \\
0 & \frac{3}{7} & 0 & 0 \\
0 & 0 & 0 & 0 \\
0 & 0 & 0 & -\frac{5}{7}
		\end{array}
		\right)$ &
		$\left(
		\begin{array}{cccc}
		\frac{2}{3} & 0 & 0 & -\frac{2}{3}\sqrt{\frac{2}{15}} \\
0 & 0 & 0 & 0 \\
0 & 0 & \frac{1}{5} & 0 \\
-\frac{2}{3}\sqrt{\frac{2}{15}} & 0 & 0 & -\frac{26}{45}
		\end{array}
		\right)$&
		$\left(
		\begin{array}{cc}
			\frac{3}{5} & 0 \\
0 & -\frac{1}{5}
		\end{array}
		\right)$
		\\	
   \hline
     $DK^*\to{}DK^*$ &  $2^{+}$ & $2^{-}$   &      $3^{+}$  &       $3^{-}$\\
     ${\cal{Z}}$&	$^{3}D_J$ & ($^{3}P_{J},^{3}F_{J}$) &$^{3}D_{J}$ &$^{3}F_{J}$
\\ $\vec{\epsilon}_{2}\cdot{}\vec{\epsilon}_{4}^{\dagger}$
 &$\left(
			\begin{array}{c}
			  1
			\end{array}
			\right)$
&$\left(
			\begin{array}{cc}
			  1&0\\0&1
			\end{array}
			\right)$
&$\left(
			\begin{array}{c}
			  1
			\end{array}
			\right)$
&$\left(
			\begin{array}{c}
			  1
			\end{array}
			\right)$\\
\hline\hline

	\end{tabular}
\end{table*}

\section{RESULTS AND DISCUSSIONS}\label{Sec: results}
\begin{table*}[htbp!]
\centering
\setlength{\tabcolsep}{9pt}
\caption{Possible bound states for the $D^{(*)}K^{*}$ interaction with $\Lambda_i = m_i + 220 \alpha~\mathrm{MeV}$ and different spin-parity assignments.
$E$ denote the eigenvalues . The symbol $\times$ indicates that no bound-state solution is found. Notation: $P$ denotes the probability (\%), while the pure-state
contributions are not shown.  Here we keep only two significant digits, which leads to the appearance of zero components.}
\label{tab:boundstates}
\begin{tabular}{ccccccccccccccccccccc}
\hline\hline
System         &State          &$\alpha$       &$E$ (MeV)      &   $P$   &State          &$\alpha$       &$E$ (MeV)             & $P$    \\\hline
$DK^*$         &$J^P=0^-$      &$1.8$          &$-0.06$        &               &$J^P=1^{-}$    &$1.8$          &$-0.06$       &              \\
               &               &$2.3$          &$-4.62$        &               &               &$2.3$          &$-4.62$       &              \\
               &               &$2.8$          &$-12.01$       &               &               &$2.8$          &$-12.01$      &              \\
               &$J^P=2^+$      &$4.2$          &$-0.54$        &               &$J^P=3^{+}$    &$4.2$          &$-0.54$       &              \\
               &               &$4.7$          &$-4.53$        &               &               &$4.7$          &$-4.53$       &              \\
               &               &$5.2$          &$-9.28$        &               &               &$5.2$          &$-9.28$       &              \\
               &$J^P=3^{-}$    &$\times$       &$\times $         \\
               &~~             &$\times$       &$\times$         \\
               &~~             &$\times$       &$\times$        \\

               &               &               &              &$P[{}^3S_{1}/{}^3D_{1}]$                 &          &       & &$P[{}^3P_{2}/{}^3F_{2}]$             \\
            &$J^P=1^+$    &$0.7$       &$-0.11$  &$100/0$            &$J^P=2^{-}$  &$1.8$          &$-0.06$  &$100/0$    \\
                            &     &$1.2$       &$-4.60$  &$100/0$        &          &$2.3$       &$-4.62$  &$100/0$ \\
                            &     &$1.7$       &$-14.56$ &$100/0$        &               &$2.8$          &$-12.01$ &$100/0$         \\ \hline \hline
         &          &       & &$P[{}^1S_{0}/{}^5D_{0}]$                 &          &       & &$P[{}^3S_{1}/{}^3D_{1}/{}^5D_{1}]$             \\
$D^*K^*$       &$J^P=0^{+}$    &$0.5$       &$-0.21$  &$99.96/0.04$                  &$J^P=1^+$    &$0.5$       &$-0.59$  &$99.83/0.17/0$       \\
                &               &$0.8$       &$-4.30$  &$99.99/0.01$              &               &$0.7$       &$-3.96$  &$99.96/0.04/0$        \\
                &               &$1.1$       &$-14.99$ &$100/0$               &               &$0.9$       &$-11.22$ &$99.99/0.01/0$      \\
         &          &       & &$P[{}^1P_{1}/{}^3P_{1}/{}^5P_{1}/{}^5F_{1}]$                 &          &       & &$P[{}^1D_{2}/{}^3D_{2}/{}^5S_{2}/{}^5D_{2}]$             \\
      &$J^P=1^{-}$  &$1.2$          &$-1.58$  &$83.29/0/16.66/0.05$            &$J^P=2^{+}$&$0.6$         &$-0.38$  &$0.25/0/99.72/0.03$           \\
                &      &$1.4$          &$-7.04$  &$83.32/0/16.66/0.01$           &~~&$0.8$         &$-2.65$  &$0.07/0/99.92/0.01$            \\
                &               &$1.6$          &$-15.52$ &$83.33/0/16.67/0.01$           &~~&$1.0$         &$-7.29$  &$0.03/0/99.97/0$        \\
         &          &       & &$P[{}^3P_{2}/{}^3F_{2}/{}^5P_{2}/{}^5F_{2}]$                 &          &       & &$P[{}^3D_{3}/{}^5D_{3}]$             \\
   &$J^P=2^{-}$&$1.2$         &$-0.43$  &$99.56/0.44/0/0$     &$J^P=3^{+}$  &$2.3$          &$-1.07$  &$100/0$                    \\
              &~~&$1.4$         &$-4.50$  &$99.86/0.14/0/0$        &               &$2.5$          &$-6.69$  &$100/0$              \\
             &~~&$1.6$         &$-11.10$ &$99.94/0.06/0/0$      &               &$2.7$          &$-14.19$ &$100/0$                   \\
 &          &               &                           &$P[{}^1F_{3}/{}^3F_{3}/{}^5P_{3}/{}^5F_{3}]$   &               &             &       &                \\
            &$J^P=3^{-}$    &$1.6$            &$-0.82$  &$0.49/0/99.45/0.05$                            &$J^P=0^{-}$    & 0.6         &-0.22                 \\
                            &~~&$1.8$         &$-3.47$  &$0.30/0/99.67/0.03$                            &               & 1.1         &-8.70      &     \\
                            &~~&$2.0$         &$-7.16$  &$0.21/0/99.76/0.02$                            &               & 1.6         &-29.87        &   \\
\hline
\hline
\end{tabular}
\end{table*}

Now, we present the numerical results in detail, as summarized in Table~\ref{tab:boundstates}. In the table, $\alpha$ is a parameter associated with the form factor, which cannot be determined from
first principles and is typically constrained by fitting to experimental data. Indeed, a wide range of values for $\alpha$, roughly spanning $0.5$--$9.7$, has been suggested in the literature through
analyses of various reaction processes. For instance, taking $\alpha = 0.5$--$5.0$ in the relation $\Lambda = m + 220\,\alpha$ provides a reasonable description of the decay widths of the hidden-charm
mesons $X(3872)\to J/\psi\,\rho$ and $Y(3940)\to J/\psi\,\omega$~\cite{Liu:2006df}, and a similar range has also been employed in other works~\cite{Colangelo:2002mj,Meng:2007cx}. By fitting the
$e^+e^- \to \bar{p}p\pi^0$ data at $\sqrt{s} = 3.773~\text{GeV}$~\cite{BESIII:2014lpf}, the parameter $\alpha$ was further constrained to be $\alpha = 6.2 \pm 3.5$~\cite{Xu:2015qqa}. Moreover, with
$\alpha$ varying from $0.53$ to $1.20$, the experimental branching ratios of $\psi(4040)\to J/\psi\,\eta$ and $\psi(4160)\to J/\psi\,\eta$ can also be well reproduced~\cite{Chen:2012nva}. Therefore,
in this work we explore possible bound states by varying $\alpha$ in the range $0.5$--$9.7$.

First, we investigate whether the $DK^{*}$ interaction alone can form a molecular state. The spin-parity quantum numbers of the system considered in this work are listed in Table~\ref{tab:spins}.
Since the $D$ and $K^{*}$ mesons have $J^P=0^{-}$ and $1^{-}$, respectively, a $DK^{*}$ molecular state with $J^P=0^{+}$ is not allowed. In the table, the notation $-$ is used to indicate the absence
of the corresponding configuration.  Compared with the non-existent $J^P=0^{+}$ $DK^{*}$ molecular state, the system can form a $J^P=0^{-}$ molecular state, where the relative orbital angular momentum
is $l=1$, leading to a relatively strong centrifugal barrier that hinders the formation of the bound state. In this case, it corresponds to a $P$-wave molecular state, and our calculation shows that only
the ${}^{3}P_{0}$ partial wave contributes.  In detail, when $\alpha = 1.8$, which lies within the range considered in this work, the $DK^{*}$ interaction
potential is just sufficient to overcome the centrifugal barrier and form a bound state, with a binding energy of $E = 0.06~\text{MeV}$.   Moreover, the binding energies of the bound states generated from the
$DK^{*}$ interaction increase with the cutoff parameter $\alpha$. For instance, when $\alpha = 2.3$, the binding energy reaches $E = 4.62~\text{MeV}$.  This implies that, as $\alpha$ increases from small to
large values, the bound state becomes increasingly deeply bound.

Additionally, we find the presence of another $P$-wave bound state in the $J^P=1^{-}$ channel, which receives contributions only from the ${}^{3}P_{1}$ partial wave. This state exhibits the same properties as the bound state appearing in the $J^P=0^{-}$ channel. Such a similarity mainly originates from the fact that both channels share the same attractive $DK^{*}$ interaction and the same centrifugal repulsion. Indeed, this conclusion can be readily drawn from Eqs.~\ref{eq10} and~\ref{eq11}. The only difference between the partial-wave contributions of the $DK^{*}$ interaction arises from the $(\epsilon_2 \cdot \epsilon_4^\dagger)$ term, whose matrix element is found to be unity for both the $J^P=0^{-}$ and $J^P=1^{-}$ channels (see Table~\ref{tab:tableqw-1}). Because the $D_{s1}(2700)$ has quantum numbers $J^{P}=1^{-}$ and its mass is about $54~\mathrm{MeV}$ below the $DK^{*}$ threshold, we find that, with $\alpha = 4.7$, the $D_{s1}(2700)$ can be reasonably interpreted within the present $P$-wave $DK^{*}$ molecular picture.

Moreover, the matrix element associated with the $(\epsilon_2 \cdot \epsilon_4^\dagger)$ term is identical in the $J^{P}=2^{+}$, $3^{+}$, and $3^{-}$ channels, taking the value of unity in all three cases. In the present analysis, these channels are dominated solely by the ${}^{3}D_{2}$, ${}^{3}D_{3}$, and ${}^{3}F_{3}$ partial waves, respectively. This feature naturally suggests similar bound-state properties for the $J^{P}=2^{+}$ and $J^{P}=3^{+}$ channels. Our numerical results indeed support this expectation: bound states in both channels first appear at $\alpha = 4.2$ with the same binding energy, $E = 0.54~\mathrm{MeV}$, and evolve almost identically as $\alpha$ increases, reaching $E = 9.28~\mathrm{MeV}$ at $\alpha = 5.2$.  By contrast, no bound state is found in the $J^{P}=3^{-}$ channel, despite the fact that it receives the same meson-exchange-induced attractive interaction as the $J^{P}=2^{+}$ and $J^{P}=3^{+}$ channels. This can be understood from the much stronger centrifugal barrier associated with the $F$-wave configuration in the $J^{P}=3^{-}$ channel, which suppresses the attractive interaction more effectively than in the corresponding $D$-wave channels.

Besides the single partial-wave contributions, the $DK^{*}$ system also exhibits coupled-channel effects in the $J^{P}=1^{+}$ and $J^{P}=2^{-}$ channels through the $^{3}S_{1}$--$^{3}D_{1}$ and $^{3}P_{2}$--$^{3}F_{2}$ mixings, respectively.  For the $J^{P}=2^{-}$ channel, a bound state emerges at $\alpha = 1.8$ with a small binding energy of $E=0.06~\mathrm{MeV}$.  As the cutoff parameter increases to $\alpha=2.3$ and $2.8$, the binding energy correspondingly rises to $4.62~\mathrm{MeV}$ and $12.01~\mathrm{MeV}$.  This behavior is fully consistent with the bound-state patterns observed in the $J^{P}=0^{-}$ and $J^{P}=1^{-}$ channels, and therefore suggests the possible existence of molecular states in the $J^{P}=0^{-}$ and $J^{P}=2^{-}$ channels with masses identical to that of the experimentally observed $D_{s1}(2700)$.  The reason is that in the $J^{P}=2^{-}$ channel, the $^{3}P_{2}$--$^{3}F_{2}$ interaction potential is decoupled (see  Table~\ref{tab:tableqw-2}) and contains only diagonal terms. Consequently, the bound-state solutions in this channel can be understood as those obtained independently from the pure $^{3}P_{2}$ and $^{3}F_{2}$ configurations. However, the $F$-wave interaction does not support a bound state, as indicated by the absence of any bound structure in the $J^{P}=3^{-}$ channel, which contains only $F$-wave contributions.  This is also the reason for the 100\% contribution of the ${}^{3}P_{2}$ partial wave to $J^{P}=2^{-}$ channel.

In contrast, in the $J^{P}=1^{+}$ channel, a bound state already emerges at the smaller cutoff value $\alpha = 0.7$, corresponding to a binding energy of $0.11~\mathrm{MeV}$. Since the one-boson-exchange model is generally expected to be most reliable for $\alpha \simeq 1$~\cite{Machleidt:1987hj}, the present results suggest that the $DK^{*}$ interaction is more likely to form a molecular state in the $J^{P}=1^{+}$ channel. It contains the feature that it is 100\% dominated by the ${}^{3}S_{1}$ partial wave, and that even when $\alpha$ is increased to $4.2$, a bound-state solution begins to appear in the pure $D$ wave at this value of $\alpha$, with its binding energy as small as $0.54~\mathrm{MeV}$.

We now turn to the analysis of the pure $D^{*}K^{*}$ system. Compared with the $DK^{*}$ case, the molecular structures generated by the $D^{*}K^{*}$ interaction are more complex, involving more partial waves for a given spin--parity configuration, as summarized in Table~\ref{tab:spins}. As in the $DK^{*}$ system, we investigate the possibility of bound states in the $S$ wave and also systematically search for bound states in higher partial waves. The corresponding numerical results are presented in Table~\ref{tab:boundstates}.

For the $J^{P}=0^{+}$ channel, which receives coupled-channel contributions from the $^1S_0$ and $^5D_0$ partial waves, a bound state emerges at $\alpha=0.5$ with a binding energy of $E=0.21~\mathrm{MeV}$. As $\alpha$ increases to $0.8$ and $1.1$, the binding energy rises to $4.30~\mathrm{MeV}$ and $14.99~\mathrm{MeV}$, respectively.  We also find that the $S$-wave component is strongly dominant, contributing about $99.96\%$ at $\alpha=0.5$, and increasing to nearly $100\%$ as $\alpha$ grows to $1.1$.  Another $D^{*}K^{*}$ molecular state with a dominant $S$-wave component is found in the $J^{P}=1^{+}$ and $J^{P}=2^{+}$ channels, where the $S$-wave fraction exceeds $99\%$. Both channels can already develop bound states at relatively small values of $\alpha$.

In contrast, the $J^{P}=0^{-}$ channel receives contributions solely from the $^{3}P_{0}$ partial wave. Due to the suppression from the $P$-wave centrifugal barrier relative to the $S$- wave, the formation of a bound state requires a stronger meson-exchange attraction, which is typically reflected by a larger cutoff parameter $\alpha$. Consequently, one expects the molecular state in the $J^{P}=0^{-}$ channel to emerge at a larger $\alpha$ than that in the $J^{P}=0^{+}$ and $J^{P}=1^{+}$ channels, where the latter is dominated by the $S$-wave component. Indeed, the $^{3}P_{0}$ $D^{*}K^{*}$ molecular state starts to appear at $\alpha=0.6$, with a corresponding binding energy of $E=0.22~\mathrm{MeV}$.

\begin{figure}[http]
\begin{center}
\includegraphics[ width=8cm]{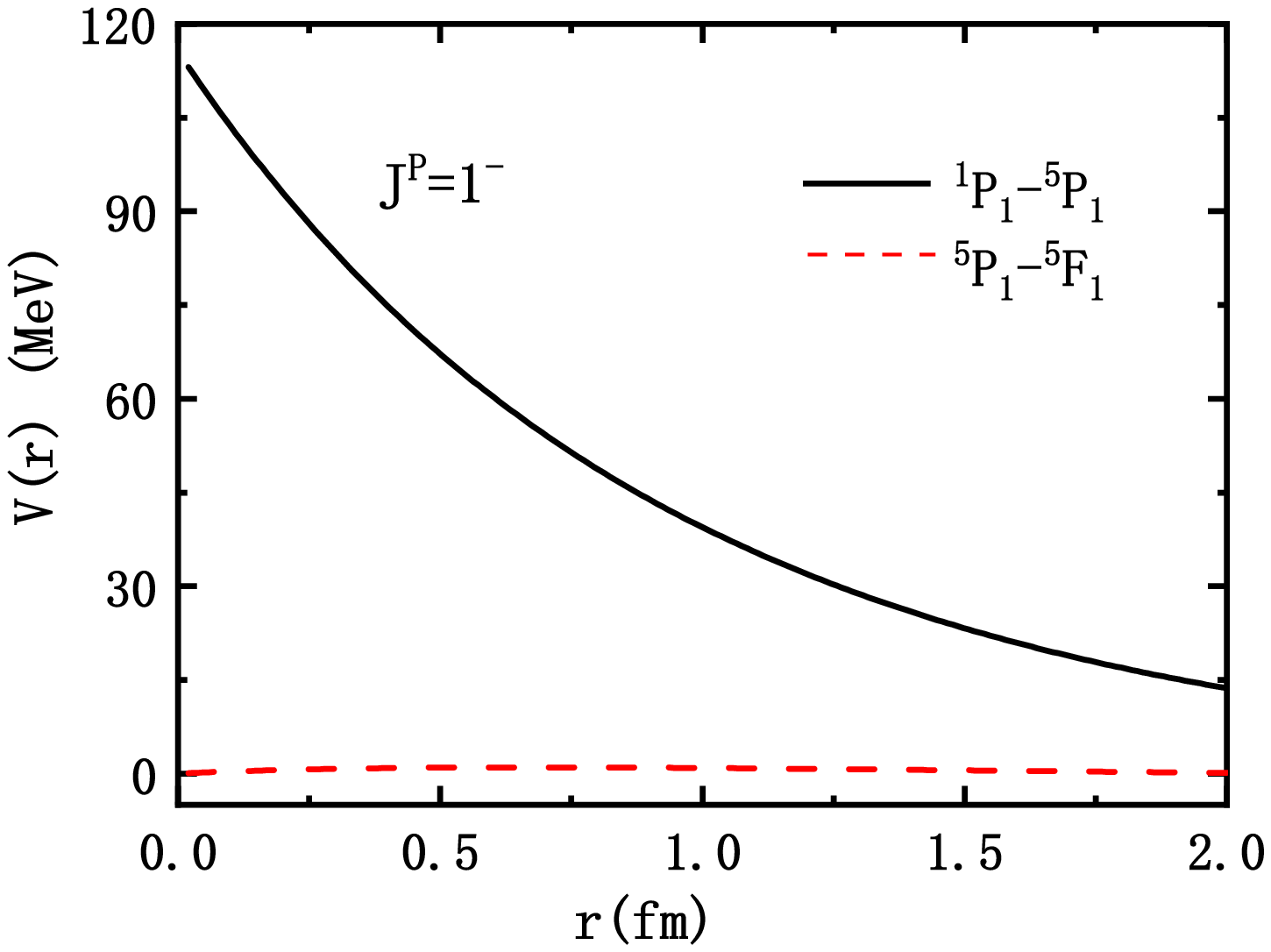}
\caption{The presence of a repulsive interaction in the $J^{P}=1^{-}$ $D^{*}K^{*}$ channel. The $^{1}P_1$--$^{5}P_1$ term denotes the coupled interaction between the $^{1}P_1$ and $^{5}P_1$ partial waves, with analogous definitions for the remaining cases. }\label{fig:dsks1-}
\end{center}
\end{figure}

For the $J^{P}=1^{-}$ channel, where both $P$- and $F$-wave contributions are included, a bound state appears at $\alpha=1.2$ with a small binding energy of $E=1.58~\mathrm{MeV}$. Compared with the $J^{P}=0^{-}$ channel, a significantly larger cutoff parameter is required to obtain a similar binding energy. For example, reproducing a state with $E=39.52~\mathrm{MeV}$ requires $\alpha=1.97$ in the $J^{P}=1^{-}$ channel, while only $\alpha=1.76$ is needed in the $J^{P}=0^{-}$ channel. A possible explanation is that the $J^{P}=0^{-}$ channel contains only the pure $^{3}P_{0}$ contribution, whereas the $J^{P}=1^{-}$ channel involves additional coupled-channel effects, including an $F$-wave component and coupled singlet-, triplet-, and quintet-spin $P$-wave contributions. These effects may reduce the effective attraction of the system. Indeed, two of the coupled-channel interaction potentials shown in Fig.~\ref{fig:dsks1-} are repulsive, which likely explains why a larger value of $\alpha$ is required to generate a bound state in the $J^{P}=1^{-}$ channel.

Remarkably, for $\alpha=1.97$, the binding energy reaches $E=39.52~\mathrm{MeV}$, yielding a molecular-state mass consistent with the experimentally observed $D_{s1}(2860)$. Therefore, our results support interpreting the $D_{s1}(2860)$ as a predominantly $P$-wave $D^{*}K^{*}$ molecular state.  Furthermore, within the same framework, the other excited $D_s$ state near $2860~\mathrm{MeV}$ with quantum numbers $J^{P}=3^{-}$ can also be accommodated as a $D^{*}K^{*}$ molecular state. The corresponding bound-state solution is overwhelmingly dominated by the $P$-wave configuration, whose fraction in the total wave function exceeds $99\%$. Therefore, our results provide a unified molecular interpretation of both excited $D_s$ states observed around $2860~\mathrm{MeV}$.  For the remaining $J^{P}=2^{-}$ and $J^{P}=3^{+}$ channels, our calculations also indicate the existence of bound states, which can be interpreted as $D^{*}K^{*}$ molecular states dominated by the $^{3}P_{2}$ and $^{3}D_{3}$ partial waves, respectively.

It is worth noting that some studies have suggested that $D_{s1}(2860)$ may correspond to an $S$-wave $D_1K$ molecular state\cite{Guo:2017jvc,Wang:2025jcq,Guo:2011dd}. In the present work, we do not include the coupling 
to this channel, mainly because we interpret $D_{s1}(2860)$ as a $P$-wave $D^{*}K^{*}$ molecular configuration, and the two states carry different quantum numbers, so no mixing between them is expected.  More importantly, 
a key motivation for interpreting $D_{s1}(2860)$
as a $D^{*}K^{*}$ molecular state is that this assignment can provide a more constrained input for symmetry-breaking effects within the heavy-quark flavor symmetry (HQFS) framework. The underlying reason is that its bottom-sector partner may be identified with the experimentally observed state $B_{sJ}(6114)$. However, no clear experimental evidence for a $B_1K$ molecular structure has been observed so far, which makes the $D^{*}K^{*}$ interpretation more advantageous in constructing testable HQFS relations.

\section{Summary}\label{sec:summary}
In the charm-strange meson sector, the $D_{s0}(2317)$ and $D_{s1}(2460)$ have been widely regarded as exotic states, since their masses lie significantly below quark model predictions and they were first observed
in the isospin-violating decay modes $\pi D_s$ and $\pi D_s^*$. At present, they are commonly interpreted as $KD$ and $KD^*$ molecular states.  Based on heavy quark flavor symmetry (HQFS), their bottom-strange
partners $B_{s0}$ and $B_{s1}$ are predicted to exist as $K\bar{B}$ and $K\bar{B}^*$ molecular states. However, these states have not yet been observed experimentally. This discrepancy between theory and experiment
suggests that the breaking mechanisms of heavy quark symmetry across different flavor sectors are not yet fully understood, primarily due to the finite mass of the heavy quark.  A natural way to address this issue
is to explore additional $K^{(*)}D^{(*)}$ and $K^{(*)}\bar{B}^{(*)}$ molecular states as experimental inputs, which may help constrain the symmetry-breaking effects and provide theoretical guidance for the yet-unobserved $K\bar{B}$ and $K\bar{B}^*$ molecular states.

In this work, we systematically investigate whether the $DK^*$ and $D^*K^*$ interactions can form molecular states within the one-boson-exchange framework.  Within the Gaussian expansion method, we solve the Schr\"{o}dinger equation with the meson-exchange potential $V(r)$ to obtain the binding energies of the molecular states.  For the $DK^{*}$ interaction, the exchanges of $\sigma$, $\rho$, and $\omega$ mesons are taken into
account, while for the $D^{*}K^{*}$ interaction, additional contributions from $\eta$ and $\pi$ exchanges are included. In the calculations, we consistently consider the contributions from $S$, $P$, $D$, and $F$
partial waves.

Our results show that the $DK^{*}$ interaction can generate bound states in the $J^{P}=0^{-},\,1^{-},\,2^{+},\,3^{+}$ channels, whereas the $D^{*}K^{*}$ interaction can form molecular states in all the spin-parity channels considered in this work. In particular, the $D_{s1}(2700)$ can be interpreted as a pure $P$-wave $DK^{*}$ molecular state, while the $D_{s1}(2860)$ and $D_{s3}(2860)$ can both be described as $D^{*}K^{*}$ molecular states, dominated by the $^{1}P_{1}$ and $^{5}P_{3}$ partial waves, respectively.  These findings provide a new theoretical perspective for understanding exotic states in the charm-strange meson spectrum. If confirmed experimentally, they may provide quantitative benchmarks for studying symmetry-breaking effects in the heavy-quark flavor sector.

\section*{Acknowledgments}
This work was supported by the National Natural Science Foundation
of China under Grant No.12005177.  Yin. Huang also acknowledges the support from the Fundamental Re
search Funds for the Central Universities under Grant No.
2682026TPY011.

%
\end{document}